\documentclass[12pt]{iopart}
\usepackage{graphicx}

\begin{document}

\title[]{Qudits of composite dimension, mutually unbiased bases and projective ring geometry}

\author{Michel Planat and Anne-C\'{e}line Baboin}

\address{Institut FEMTO-ST, CNRS, D\' epartement LPMO, 32 Avenue de
l'Observatoire\\ F-25044 Besan\c con, France }
\ead{michel.planat@femto-st.fr}
\begin{abstract}
The $d^2$ Pauli operators attached to a composite qudit in dimension $d$ may be mapped to the vectors of the symplectic module $\mathcal{Z}_d^{2}$ ($\mathcal{Z}_d$ being the modular ring). As a result, perpendicular vectors correspond to commuting operators, a free cyclic submodule to a maximal commuting set, and disjoint such sets to mutually unbiased bases. For dimensions $d=6,~10,~15,~12$, and $18$, the fine structure and the incidence between maximal commuting sets is found to reproduce the projective line over the rings $\mathcal{Z}_{6}$, $\mathcal{Z}_{10}$, $\mathcal{Z}_{15}$, $\mathcal{Z}_6 \times \mathbf{F}_4$ and $\mathcal{Z}_6 \times \mathcal{Z}_3$, respectively. 
\end{abstract}

\maketitle

%\noindent
%{\bf PACS Numbers:} 03.67.-a, 03.65.Fd, 02.10.Ox, 02.40.Dr\\
%{\bf Keywords:} Generalized Pauli Operators --- Mutually Unbiased Bases
%--- Finite Rings-- Projective Geometries.

\noindent
\hrulefill
\section*{Summary}

Commutation relations of (generalized) Pauli operators provide a skeleton for mutually unbiased bases, quantum entanglement and other conceptual (or practical) issues like quantum computing \cite{sigma,Pauligraphs}. Recently, an extensive study of commuting/non commuting rules has been undertaken, firstly in prime power dimensions $d=p^N$ of the Hilbert space \cite{sigma}-\cite{Havlicek}, then in the smallest composite dimension $d=6$ \cite {IJTP08}. Commutation relations of two-qubit operators, and dually the incidence relations between maximal commuting sets of them, have been shown to fit the (symplectic) generalized quadrangle of order two, and several projective embeddings have been proposed \cite{Pauligraphs,Veldkamp}. For higher-order Pauli operators, the duality between the observables and their maximal commuting sets does not occur and the geometrical space of points/observables may exhibit several lines/sets passing through $n$-tuples of distinguished points \cite{IJTP08}.

In this communication, one makes use of the maximal ideals of some ring $\mathcal{R}$ (possibly different from the modular ring $\mathcal{Z}_d$) as the gears of commutation relations. 
%These unusual findings may pave the way for future applications in quantum information processing.
In particular, the incidence between the twelve lines of the sextit system fits the grid like structure of the projective line $P_1(\mathcal{Z}_6)$ over the modular ring $\mathcal{Z}_6$.
%, but the multiline projective structure of the $d^2=36$ Pauli observables (including the unity matrix) has remained elusive\cite{IJTP08}.
In the higher composite dimensions explored so far $d=2\times5=10$, $d=3\times 5=15$, $d=2 \times 3^2=18$ and $d=2^2 \times 3=12$, the incidence of the maximal commuting sets is found to reproduce the projective line $P_1(\mathcal{R})$ over rings $\mathcal{R}=\mathcal{Z}_{10}$, $\mathcal{R}=\mathcal{Z}_{15}$, $\mathcal{R}=\mathcal{Z}_6 \times \mathcal{Z}_3$ and $\mathcal{R}=\mathcal{Z}_6 \times \mathbf{F}_4$, respectively. The unexpected irruption of the Galois field of four elements $\mathbf{F}_4$, within the projective model of the two-qubit/qutrit system, seems to forbid an easy generalization to an arbitrary dimension $d$.

There are indeed many ways of defining the quantum states (let us call them \lq\lq qudits") in a finite $d$-dimensional Hilbert space. One representation makes use of the unitary \lq\lq shift" and \lq\lq clock" operators $X$ and $Z$, with the actions $X \left|s\right\rangle=\left|s+1\right\rangle$,  $Z \left|s\right\rangle=\omega^s\left|s\right\rangle$ on the vectors $\left|s\right\rangle$ of the Hilbert space. Henceforth $\omega$ is a fixed $d$-th root of the unity. Under matrix multiplication, $X$ and $Z$ generates the (non-commutative) Pauli group $G$ from the basic relation $Z X=\omega X Z$. As a result, elements of $G$ can be taken as $\omega^a X^b Z^c$, with $a$, $b$ and $c$ in the ring $\mathcal{Z}_d$ \cite{Vourdas}-\cite{HansH}. 
Another representation of the Pauli group is from tensor products of shift and clock actions in prime dimension \cite{Gottesman,Bandyo}. 
%As soon as one removes the phase factor $\omega^a$ in the generic elements of $G$, one arrives at the $d^2$ Pauli operators (including unity matrix).
The latter definition is favored in the theory of mutually unbiased bases \cite{Bandyo,EPJD} and was used in our previous papers devoted to commutation relations \cite{sigma}-\cite{IJTP08}. A condensation from the $d^3$ elements of the Pauli group to $d^2$ Pauli operators  may also be achieved by taking the quotient of $G$ by its center $G'$ (the set of all operators which commute with every other one)\footnote{See \cite{Howe} for a deep connection between mutually unbiased bases and the maximal isotropic subspaces attached to the finite Heisenberg group over a ring, and also \cite{PlanatRH} for an intriguing connection of phase-locked quantum states to prime number theory and the Riemann hypothesis.} 

Ref \cite{HansH} describes the commutation relations between operators in $G$, and thus in $G/G'$, using vectors $(b,c)\in \mathcal{Z}_d^2$, their attached cyclic submodule
\begin{equation}
\mathcal{Z}_d(b,c)=\{(ub,uc):u \in \mathcal{Z}_d\},
\end{equation}        
and the \lq\lq points" of the projective line
\begin{equation}
\mathcal{P}_1(\mathcal{Z}_d)=\{\mathcal{Z}_d(b,c): (b,c)~ \mbox{is}~ \mbox{admissible}\}.
\end{equation}        
An admissible vector $(b,c)$ is such that there exists another vector $(x,y)$ for which the matrix $\left(
\begin{array}{cc}
b & c \\
x & y \\
\end{array}
\right)$ is invertible, which for a commutative ring is equivalent to have a determinant equal to a unit of the ring. The equivalence class of $(b,c)$ is a free cyclic submodule $\mathcal{Z}_d(b,c)$, of order $d$, and also a \lq\lq point" of the projective line $\mathcal{P}_1(\mathcal{Z}_d)$.

One reminds the geometrical structure of the projective line $\mathcal{P}_1(\mathcal{Z}_d)$ \cite{Havl00,projlines}. Two distinct points $\mathcal{Z}_d(b,c)$ and $\mathcal{Z}_d(b',c')$ are called distant if $\det\left(\begin{array}{cc}
b & c \\
b' & c'\\
\end{array}
\right)$ 
equals a unit of the ring $\mathcal{Z}_d$. Otherwise the two points belong to the same neighborhood. 

Another crucial concept organizes the vectors in $\mathcal{Z}_d^2 $: a perpendicular set $(b,c)^\perp$ is defined as 
\begin{equation}
(b,c)^\perp=\{(u,v) \in \mathcal{Z}_d^{2} :(b,c)\perp(u,v)\},
\end{equation}        
in which two vectors $(b,c)$ and $(u,v)$ are perpendicular if $\det\left(\begin{array}{cc}
b & c \\
u & v \\
\end{array}
\right)=0$. Note that two vectors within a cyclic submodule are mutually perpendicular. According to \cite{HansH}, operators in $G$ which commute with a fixed operator correspond to a perpendicular set\footnote{This notion of perpendicularity related to the commutativity of the operators was already used within the context of symplectic polar spaces as models of $N$-qubit systems (see \cite{Saniga1} and Sec 4.1 of \cite{Pauligraphs}).}. Using this analogy, it seems natural to identify the elements of a free cyclic submodule, which are mutually perpendicular, with the maximal commuting sets of Pauli operators, as we already did it implicitely in \cite{IJTP08}. A posteriori one should not be surprised that the projective line $\mathcal{P}_1(\mathcal{Z}_6)$ fits the incidence relations between the maximal commuting sets of the sextit system. To complete the geometrical picture of commutation relations, one needs to identify the (not necessarily admissible) vectors of  $\mathcal{Z}_d^{2}$ with the $d^2$ Pauli operators. 

Let us summarize main results of \cite{HansH}:

 Theorem 1 asserts that a free cyclic module $\mathcal{Z}_d(b',c')$ containing a vector $(b,c)$ is contained in the perpendicular set $(b,c)^\perp$. Only if $(b,c)$ is admissible the corresponding module equals $(b,c)^\perp$.
 
It reinforces our interpretation that the maximal sets of mutually commuting operators [corresponding to $\mathcal{Z}_d(b,c)$] also define a base of operators [corresponding to $(b,c)^\perp$].

 One immediate consequence concerns the application to mutually unbiased bases. Any two vectors in one base should be perpendicular, while any two vectors from distinct mutually unbiased bases should not. Using two non-zero (and admissible) distinct vectors $(b,c)$ and $(b',c')$, the two vector sets
$\mathcal{Z}_d(b,c)\setminus \{(0,0)\}=\{(ub,uc):u \in \mathcal{Z}_d \setminus \{0\}\}$
 and $\mathcal{Z}_d(b',c')\setminus \{(0,0)\}=\{(vb',vc'):v \in \mathcal{Z}_d \setminus \{0\}\}$  are disjoint only if $uv(bc'-cb') \neq 0$, i.e. if $uv \neq 0$ and $(b,c)$, $(b',c')$ are not perpendicular. This cannot happen maximally since $\mathcal{Z}_d$ is a ring so that $u$ or $v$ may be zero divisors. 
The maximal number of mutually unbiased bases in composite dimension may thus be reformulated as being the maximal number of such disjoint vector sets in the relevant ring.

If the dimension $d$ is the power of distinct primes $p_k$, theorem 2 in \cite{HansH} provides quantitative results about (a) the number of points $n_d$ in which any vector $(b,c)$ lies, (b) the partitioning of $(b,c)^\perp$ as the corresponding set theoretic union of points $\mathcal{Z}_d(b,c)$ and (c) the cardinality of $(b,c)^\perp$. One gets 

\begin{equation}
n_d=\prod_{k\in K} (p_k+1)~~\mbox{and}~~ |(b,c)^\perp|=d\prod_{k\in K} p_k,
\label{quant}
\end{equation}
in which $K$ is a subset of the indices related to the decomposition of the entries  of $(b,c)$ into their principal ideals.

\section*{Commutation relations of the sextit system}

The sextit system ($d=2\times3=6$) was investigated in our recent paper \cite{IJTP08}. In this dimension, the (generalized) Pauli operators are defined as
\begin{equation}
\sigma_i \otimes \sigma_j,~~i\in
\{0,\ldots,3\},~~j\in \{0,\ldots, 8\},~~(i,j)\neq(0,0).
\end{equation}
The orthonormal set of the qubits comprises the standard Pauli matrices $\sigma_i=
(I_2,\sigma_x, \sigma_y,\sigma_z)$, where
$I_2=\left(\begin{array}{cc}1 & 0 \\0 & 1\\\end{array}\right)$,
$\sigma_x=\left(\begin{array}{cc}0 & 1 \\1 &
0\\\end{array}\right)$, $\sigma_z=\left(\begin{array}{cc}1 & 0 \\0
& -1\\\end{array}\right)$ and $\sigma_y=i \sigma_x \sigma_z$, while the orthonormal set of the qutrits is taken as\\ 
$\sigma_j=\{I_3,Z,X,Y,V,Z^2,X^2,Y^2,V^2\}$,
where $I_3$ is the $3 \times 3$ unit matrix,
$Z=\left(\begin{array}{ccc}1 & 0&0
\\0 & \omega&0\\0&0& \omega^2\\\end{array}\right)$,
$X=\left(\begin{array}{ccc}0 & 0&1 \\1 & 0&0\\0&1&
0\\\end{array}\right)$, $Y=XZ$, $V=X Z^2$ and $\omega=\exp\left(2
i \pi/3\right)$.

The sextit operators can be conveniently labelled as follows: $1=I_2 \otimes
\sigma_1$, $2=I_2 \otimes \sigma_2$, $\cdots$, $8=I_2 \otimes
\sigma_8$ , $a_0=\sigma_z \otimes I_2$, $9=\sigma_z \otimes
\sigma_1$,\ldots, $b_0=\sigma_x \otimes I_2$, $17=\sigma_x \otimes
\sigma_1$,\ldots , $c_0=\sigma_y \otimes I_2$,\ldots$,32=\sigma_y
\otimes \sigma_8$, in which we singled out the three reference points $a_0$, $b_0$ and $c_0$. 

Then one can use the strategy already described in \cite{Pauligraphs} for $N$-qudit systems. The Pauli operators are identified with the vertices of a (Pauli) graph and the commuting operators are identified with the edges. The maximal cliques of the graph correspond to the maximal sets of mutually commuting operators. For the sextit system one gets the twelve sets
\footnotesize
\begin{eqnarray}
&L_1=\{1,5,a_0,9,13\},~~L_2=\{2,6,a_0,10,14\},~~L_3=\{3,7,a_0,11,15\},~~L_4=\{4,8,a_0,12,16\}, \nonumber \\
&M_1=\{1,5,b_0,17,21\},~~M_2=\{2,6,b_0,18,22\},~~M_3=\{3,7,b_0,19,23\},~~M_4=\{4,8,b_0,19,24\}, \nonumber\\
&N_1=\{1,5,c_0,25,29\},~~N_2=\{2,6,c_0,26,30\},~~N_3=\{3,7,c_0,27,31\},~~N_4=\{4,8,c_0,28,32\}\nonumber.
\end{eqnarray}
\normalsize

\begin{figure}[ht]
%\vspace*{8.0cm}
\centerline{\includegraphics[width=8.0truecm,clip=]{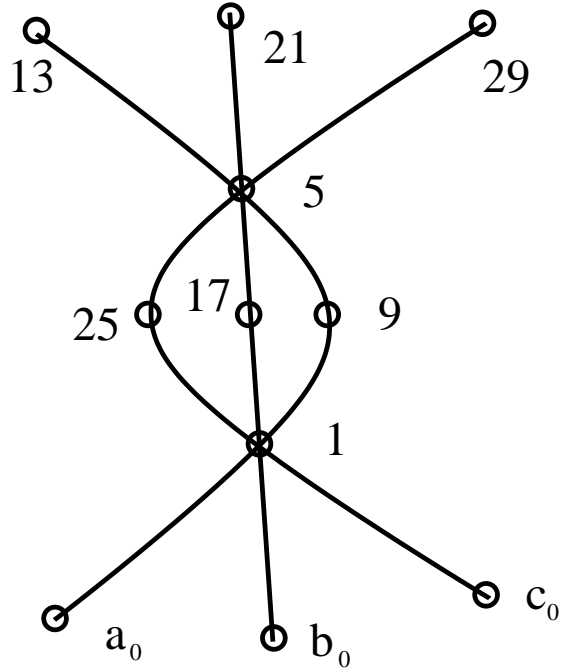}}
\caption{A sketch of a perpendicular set $x^\perp$ attached to a point of type (ii) (see the text for a definition). The whole structure comprises four similar sets having the operators $a_0$, $b_0$ and $c_0$ in common.}  \label{multi}
\end{figure}
As emphasized in \cite{IJTP08}, the incidence between the maximal commuting sets leads to a $3 \times 4$ grid-like structure isomorphic to the projective line over the ring $\mathcal{Z}_6=\mathcal{Z}_2 \times \mathcal{Z}_3$. A subset of the commutation structure of the operators is illustrated in Fig~\ref{multi}. 
\begin{figure}[ht]
%\vspace*{8.0cm}
\centerline{\includegraphics[width=12.0truecm,clip=]{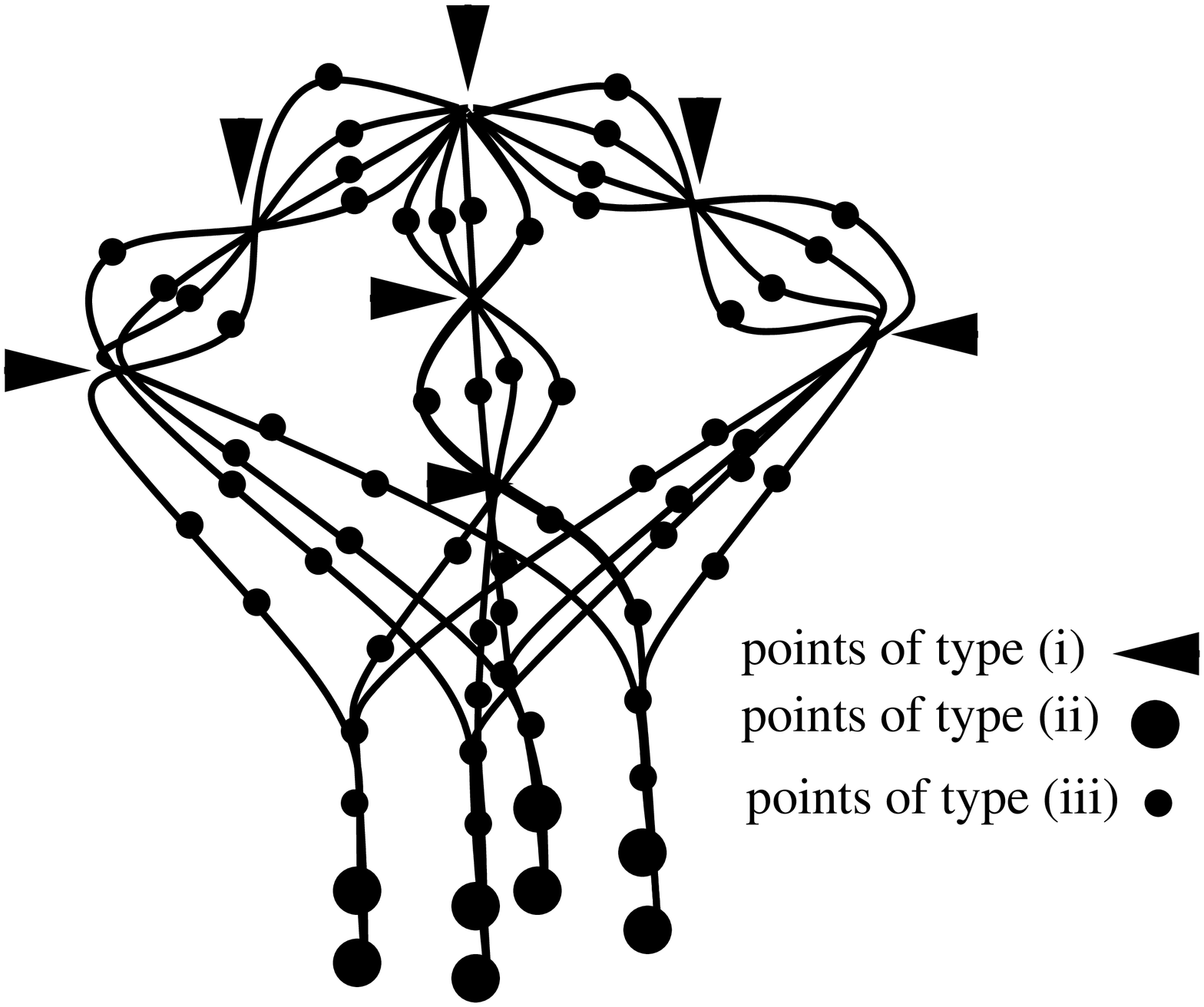}}
\caption{A sketch of a perpendicular set $x^\perp$ ($x$ is the reference point at the top of the parachute like structure. It comprises twelve maximal commuting sets, each one having eleven points (the unity operator is omitted). The three types of points (i), (ii) and (iii) are described in the text.}  \label{para}
\end{figure}
Let us illustrate the relationship between the Pauli graph of sextits and the fine structure of the projective line $P_1(\mathcal{Z}_6)$. Operators $x$ belonging to the maximal sets are of three distinct types\footnote{The perpendicular set $x^\perp$ includes the operator $x$ itself and the unity operator \cite{HansH}. But for maximal commuting sets one usually ignores the unity operator which commutes with every other operator \cite{Pauligraphs,Bandyo}} (see also Fig~\ref{multi})

(i) $x$ is one of the reference points $a_0$, $b_0$ or $c_0$, lies in four sets and the number of points commuting with $x$  is $|x^\perp|=18$,

(ii) $x\in \{1,2,3,4,5,6,7,8\}$ lies in three sets and $|x^\perp|=12$,

(iii) otherwise $x$ lies in a single set and $|x^\perp|=6$.

These results clearly fits (\ref{quant}) with $d=6$, $p_1=2$ and $p_3=3$. 

Analogous results are indeed obtained for square free dimensions $d=2\times 5=10$ and $d=3\times 5=15$, so far explored.  

\section*{Commutation relations for qudits in dimension twelve}

The qudit system in dimension $d=2^2\times3=12$ contains the even square $2^2$. In this dimension, the (generalized) Pauli operators are defined as
\begin{equation}
\sigma_i \otimes \sigma_j \otimes \sigma_k,~~i,j\in
\{0,\ldots,3\},~~k\in \{0,\ldots, 8\},~~(i,j,k)\neq(0,0,0).
\end{equation}

One proceeds as for the sextit system, one determines the Pauli graph of the $12$-dit and one extracts the maximal cliques. The incidence between the corresponding maximal commuting sets is found to reproduce\footnote{For a classification of projective lines over small commutative rings see Ref~\cite{projlines}.} the projective line over the ring $\mathcal{R}=\mathcal{Z}_{p_1}\times \mathcal{Z}_{p_2} \times \mathbf{F}_{q^2}$, of order $|\mathcal{R}|=(p_1+1)(p_2+1)(q^2+1)$ with $p_1=q=2$ and $p_2=3$. 

Operators $x$ belonging to the maximal sets still are found to be of three distinct types 

(i) $x$ is one of the reference points (it includes $I_3$ in its tensor product), then one finds that $x$ lies in $(p_1+1)(p_2+1)=12$ sets and $|x^\perp|=dp_1p_2=72$.

(ii) $x$ includes $I_2\otimes I_2$ in its tensor product, lies in $(p_1+1)(q^2+1)=15$ sets and $|x^\perp|=dp_1q=48$ ,

(iii) otherwise $x$ lies in $p_1+1=3$ sets and $|x^\perp|=p_1d=24$. 

The commutation relations within a perpendicular set $x^\perp$ of type (i) are illustrated in Fig~\ref{para}. It comprises three bundles of four lines each, organized in a parachute like structure. The lines of a specific bundle intersect at three distinguished points, each one of type (i).

\section*{Commutation relations for qudits in dimension eighteen}
\begin{figure}[ht]
%\vspace*{8.0cm}
\centerline{\includegraphics[width=12.0truecm,clip=]{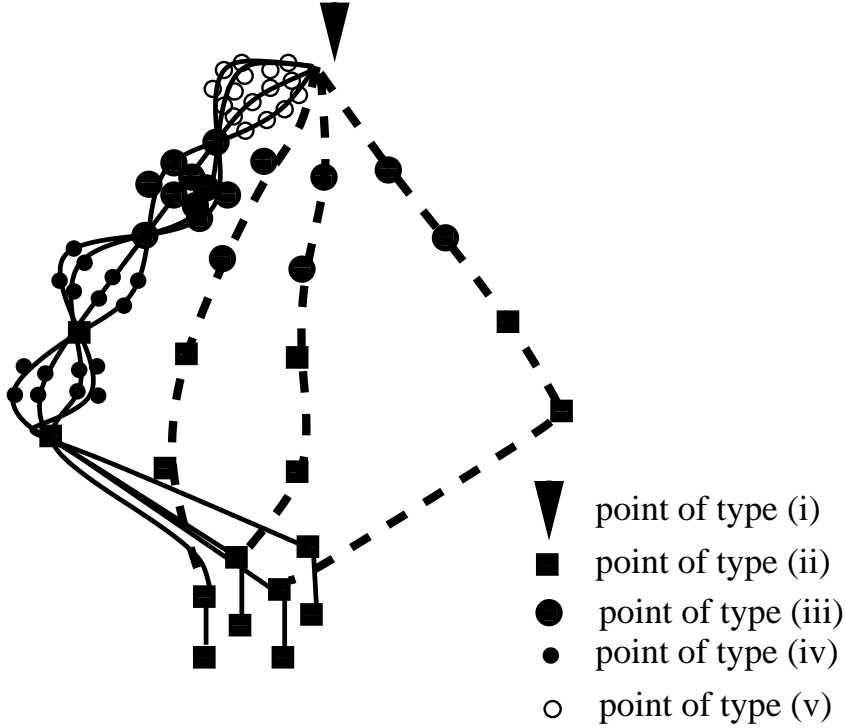}}
\caption{A sketch of a perpendicular set $x^\perp$ ($x$ is the reference point [of type (i)] at the top of the parachute like structure. It comprises sixteen maximal commuting sets, each one having seventeen points (the unity operator is omitted). Only one bundle is represented in detail. The five types of points (i) to (v) are described in the text.}  \label{paranew}
\end{figure}
The qudit system in dimension $d=2\times3^2=18$ contains the odd square $3^2$. In this dimension, the (generalized) Pauli operators are defined as
\begin{equation}
\sigma_i \otimes \sigma_j \otimes \sigma_k,~~i\in
\{0,\ldots,3\},~~j,k\in \{0,\ldots, 8\},~~(i,j,k)\neq(0,0,0).
\end{equation}

Again one determines the Pauli graph of the $18$-dit and one computes the maximal cliques. The incidence between the corresponding maximal commuting sets is found to reproduce the projective line $P_1(\mathcal{R})$ over the ring  $\mathcal{R}=\mathcal{Z}_{p_1}\times \mathcal{Z}_{p_2} \times \mathcal{Z}_{p_2}$, of order $|\mathcal{R}|=(p_1+1)(p_2+1)^2$ with $p_1=2$ and $p_2=3$.

Operators $x$ belonging to the maximal sets are found to be of five distinct types 

(i) $x$ is one of the three reference points containing $I_3\otimes I_3$ in the tensor decomposition, it lies in $(p_2+1)^2=16$ sets and $|x^\perp|=dp_2^2=162$,  

(ii) $x$ lies in $(p_1+1)(p_2+1)=12$ sets and $|x^\perp|=dp_1p_2=108$,

(iii) $x$ lies in $p_2+1=4$ sets and $|x^\perp|=dp_2=54$ ,

(iv) $x$ lies in $p_1+1=3$ sets and $|x^\perp|=d p_1 p_2=108 $,

(v) otherwise $x$ lies in a single set and $|x^\perp|=dp_2=54$. 

The perpendicular set attached to a point of type (i) is illustrated in Fig~\ref{paranew}. The fine structure of the bundles increases in complexity compared to Fig~\ref{para}, each one comprising four lines intersecting at five points, one of type (i), two of type (ii) and the remaining two of type (iii).

\section*{Discussion and conclusion}

It has been found that commuting operators associated to composite qudits in dimension $d$ correspond to perpendicular vectors within the symplectic module $\mathcal{Z}_d^{2}$. Moreover the maximal commuting sets reflect the set-theoretic structure of free cyclic submodules defined over some commutative ring $\mathcal{R}$, possibly distinct from the modular ring $\mathcal{Z}_d$ as soon as $d$ contains squares in the prime number decomposition. An admissible vector, which defines such a submodule, is of two types \cite{projlines} (a) either one (at least) of its entries is a unit of the ring $\mathcal{R}$, or (b) both of its entries are zero divisors, not in the same maximal ideal of $\mathcal{R}$. Thus the maximal ideals underlie the projective line \cite{projlines} and the commutation structure of qudit operators.
\begin{figure}[ht]
%\vspace*{8.0cm}
\centerline{\includegraphics[width=12.0truecm,clip=]{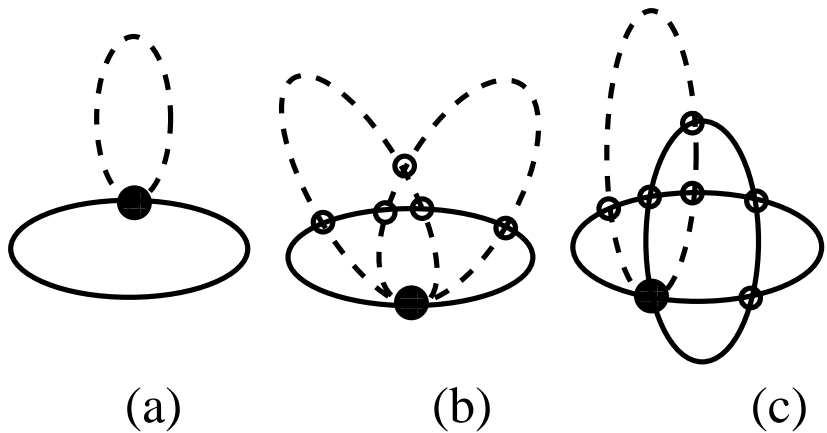}}
\caption{A sketch of the maximal ideals of rings $\mathcal{Z}_2 \times \mathcal{Z}_3$ (illustrating the qubit/qutrit)(a), $\mathcal{Z}_2 \times \mathcal{Z}_3 \times \mathcal{Z}_3$ (illustrating the two-qutrit/qubit (b) and $\mathcal{Z}_2 \times \mathcal{Z}_3 \times \mathbf{F}_4$ (illustrating the two-qubit/qutrit)(c). The ellipses feature maximal ideals, and their intersection is marked by a small circle; the filled black circle is the zero element of the ring (compare Fig~5 in \cite{sigma}).}  \label{ideals}
\end{figure}

In Fig~\ref{ideals} we give a sketch of the interaction between maximal ideals of the rings $\mathcal{Z}_2 \times \mathcal{Z}_3$ (corresponding to the qubit/qutrit system), $\mathcal{Z}_2 \times \mathcal{Z}_3 \times \mathcal{Z}_3$ (corresponding to the qubit/two-qutrit system) and $\mathcal{Z}_2 \times \mathcal{Z}_3 \times \mathbf{F}_4$ corresponding to the two-qubit/qutrit system). To some extent one can identify the factors of the qudit system with the maximal ideals, and the peculiar set theoretic union/intersection of them governs the whole commutation structure. The ideals themselves have a ring structure. For example the three ideals in (c) are subsets isomorphic to $\mathcal{Z}_2 \times \mathcal{Z}_3$,  $\mathcal{Z}_2 \times \mathbf{F}_4$ and  $\mathcal{Z}_3 \times \mathbf{F}_4$ respectively. The corresponding projective lines are $3\times 4$, $3\times 5$ and $4\times 5$ grids. The last grid exhibits a maximum number of four distant points, corresponding to the maximum number of mutually unbiased bases in dimension twelve.

Further work should clarify whether a ring $\mathcal{R}$ is attached to any composite qubit. This could have application not only to mutually unbiased bases, but to quantum chemistry \cite{Kibler}, quantum channels \cite{Nathanson}, the non abelian hidden subgroup problem \cite{Radhakrishnan} and other quantum information processing problems, as well.

\section*{Acknowledgments}

The authors acknowledge Hans Havlicek, Metod Saniga, Peter Pracna and Maurice Kibler for helpful interactions during the workshop \lq\lq Finite projective geometries in quantum theory" [http://www.ta3.sk/~msaniga/QuantGeom.htm], held in Tatranska-Lomnica in august 2007. The work was supported under the ECO-NET project 12651NJ \lq\lq Geometries over finite rings and the properties of mutually unbiased bases" and the CNRS-SAV project 20246 \lq\lq Projective and related geometries for quantum information".


\section*{Bibliography}

\vspace*{.0cm} \noindent
\vspace*{-.1cm}
\begin{thebibliography}{10}


\bibitem{sigma}
Planat M, Saniga M and  Kibler M~R 2006 Quantum entanglement and projective ring geometry {\it SIGMA} {\bf 2} Paper 066



\bibitem{Pauligraphs}
Planat M and Saniga M 2008 On the Pauli graphs of $N$-qudits {\it Quantum Information and Computation} {\bf 8} 127--146


\bibitem{Saniga1}
Saniga M and Planat M 2007 Multiple qubits as symplectic polar spaces
of order two {\it Adv. Studies Theor. Phys.} {\bf 1} 1-4 



\bibitem{Havlicek}
Havlicek H 2007 A mathematician's insight into the Saniga-Planat theorem (available on-line from http://www.geometrie.tuwien.ac.at/havlicek/talks.html)


\bibitem{IJTP08}
Planat M, Baboin A~C and Saniga M 2007 Multi-line geometry of qubit/qutrit and higher order Pauli operators {\it Preprint} 0705.2538 [quant-ph]
({\it Int. J. Theor. Phys.} accepted) 

%\bibitem{KThas}
%K. Thas (2007), {\it Pauli operators of $N$-qubit Hilbert spaces and the Saniga-Planat conjecture}, Chaos, Solitons and Fractals, accepted.


\bibitem{Veldkamp}
Saniga M, Planat M, Pracna P and Havlicek H 2007 The Veldkamp space of two-qubits {\it SIGMA} {\bf 3} Paper 075

\bibitem{Vourdas}
Vourdas A 2007 Quantum systems in finite Hilbert space: Galois fields in quantum mechanics {\it J. Phys. A: Math. Theor.} {\bf 40} R285-R331

\bibitem{Sulc}
Sulc P and Tolar J 2007 Group theoretical constructions of mutually unbiased bases in Hilbert spaces of prime dimensions {\it Preprint} 0708.4114 [quant-ph]   



\bibitem{HansH}
Havlicek H and Saniga M 2007 Projective ring line of a specific qudit {\it Preprint} 0708.4333 [quant-ph] ({\it J. Phys. A: Math. Theor.} accepted).


\bibitem{Gottesman}
Gottesman D 1998 Fault-tolerant quantum computation with higher-dimensional systems {\it Lecture Notes in Computer Science} {\bf 1509} 302-313


\bibitem{Bandyo}
Bandyopadhyay S, Boykin P~O, Roychowdhury V and Vatan F 2002 A new proof for the existence of MUBs {\it Algorithmica} {\bf 34} 512


\bibitem{EPJD}
Planat M and Rosu H~C 2005 Mutually unbiased phase states, phase uncertainties and Gauss sums {\it Eur. Phys. J D} {\bf 36} 133-139

\bibitem{Howe}
Howe 2005 R Nice error bases, mutually unbiased bases, induced representations, the Heisenberg group and finite geometries {\it Indag Mathem, N S} {\bf 16} (3-4), 553--583 

\bibitem{PlanatRH}
Planat M 2006 Huyghens, Bohr, Riemann and Galois: phase-locking {\it Int J Mod Phys B} {\bf 20} 1833-1850


\bibitem{Havl00}
Blunck A and Havlicek H 2000 Projective representations: I: Projective lines over a ring {\it Abh Math Sem Univ Hamburg} {\bf 70} 287-299

\bibitem{projlines}
Saniga M, Planat M, Kibler M~R and Pracna P 2007 A classification of the projective lines over small rings {\it Chaos, Solitons and Fractals} {\bf 33} 1095-1102


\bibitem{Kibler}
Albouy O and  Kibler M~R 2007 $SU_2$ non standard bases: the case of mutually unbiased bases {\it SIGMA} {\bf 3} Paper 076



\bibitem{Nathanson}
Nathanson M and  Ruskai M~B 2007 Pauli diagonal channels constant on axes {\it J. Phys. A: Math. Theor.} {\bf 40} 8171-8204


\bibitem{Radhakrishnan}
Radhakrishnan J, R\"{o}tteler M and Sen P 2005 On the power of random bases in Fourier sampling: hidden subgroup problem in the Heisenberg group {\it Lecture Notes in Computer Science} {\bf 3580} 1399-1411

\end{thebibliography}
\end{document}